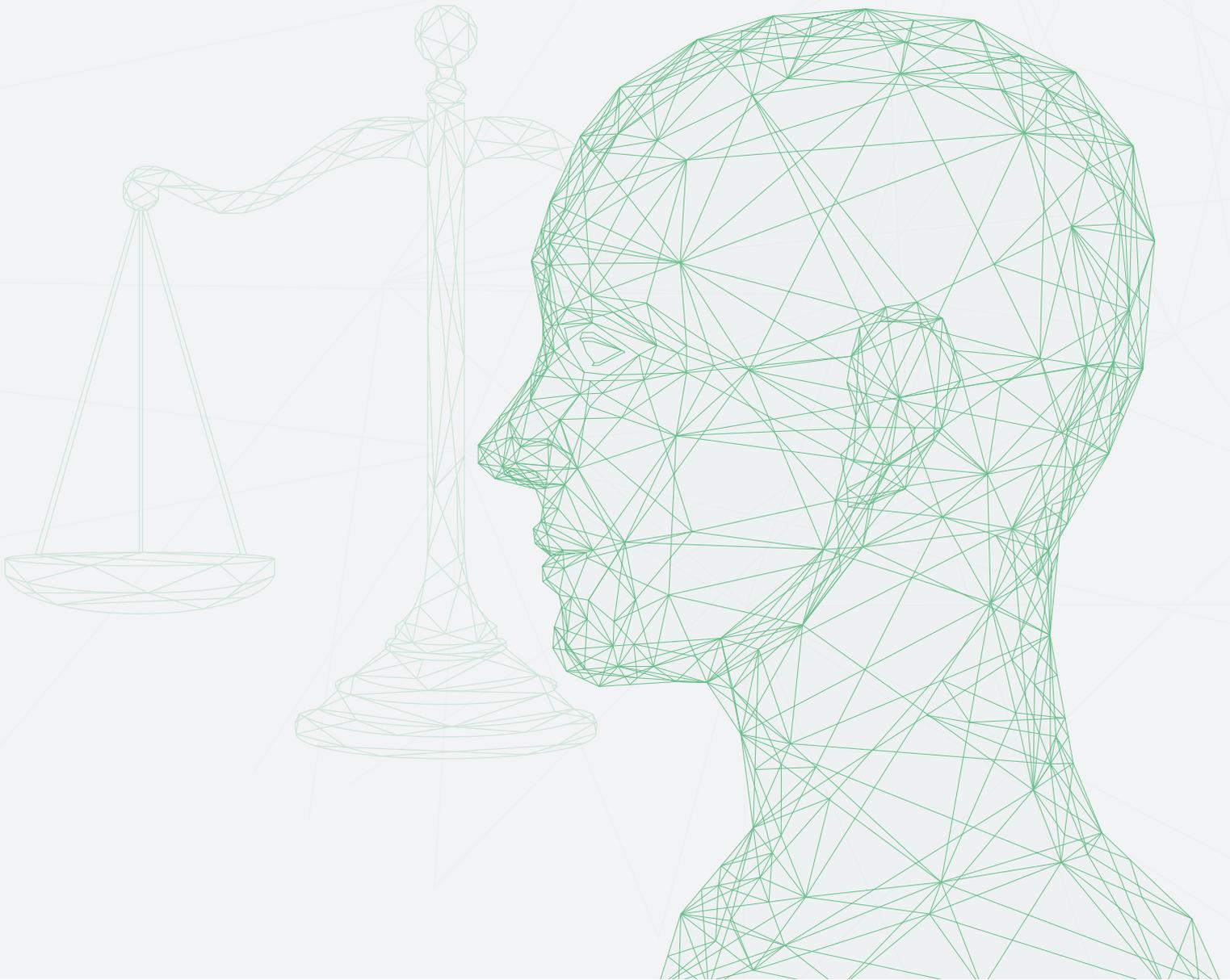

# Catalog of General Ethical Requirements for AI Certification

Nicholas Kluge Corrêa    Julia Maria Mönig

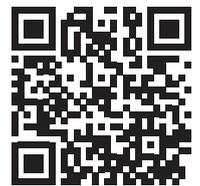

## Imprint

**Authors:**
Dr. Nicholas Kluge Corrêa kluge@uni-bonn.de
Dr. Julia Maria Mönig moenig@uni-bonn.de

Center for Science and Thought
Universität Bonn
Regina-Pacis-Weg 3
53113 Bonn
Germany
https://www.cst.uni-bonn.de/en

**Layout & Design:**
Laura Hinz

The Center for Science and Thought (CST) at the University of Bonn, Germany, is a radically interdisciplinary platform designed to address urgent questions that arise at the intersection of philosophy and various natural sciences.

„Zertifizierte KI" (Certified AI) is a KI.NRW-flagship project funded by the Ministerium für Wirtschaft, Industrie, Klimaschutz und Energie des Landes Nordrhein-Westfalen (Ministry for Economic Affairs, Industry, Climate Action and Energy of the State of North Rhine-Westphalia).

All links last accessed in October 2024.

Bonn, October 2024

# Summary

This whitepaper offers normative and practical guidance for developers of artificial intelligence (AI) systems to achieve „Trustworthy AI". In it, we present overall ethical requirements and six ethical principles with value-specific recommendations for tools to implement these principles into technology. Our value-specific recommendations address the principles of fairness, privacy and data protection, safety and robustness, sustainability, transparency and explainability and truthfulness. For each principle, we also present examples of criteria for risk assessment and categorization of AI systems and applications in line with the categories of the European Union (EU) AI Act. Our work is aimed at stakeholders who can take it as a potential blueprint to fulfill minimum ethical requirements for trustworthy AI and AI Certification.

# Contents

# 1 Introduction

Nowadays, our most recent AI revolution [114, 73, 166] has ushered in a new era of renewed interest in the field, given its success in (1) triumphing where past paradigms have stagnated (e.g., object detection, language understanding, etc.) and (2) offering new technologies for the industry to adopt into its practices. For example, recent developments in the field of generative AI have unleashed upon the world systems that can ingest and generate content, like text, images, and video, with a remarkable level of resemblance to what human beings can produce, integrating AI increasingly into many parts of modern-day life.

While the true nature and extent of the capabilities of such systems remain a matter of debate and speculation [26, 103, 144], these same capacities have sparked several worries regarding the type of future we are creating, where intelligent systems become an ever-growing part of society. From pleas for moratoriums [67] to proposals of general research agendas to map and mitigate the risks associated with AI [18], society is starting to create a consensus on the need to imbue artificial intelligence with values and norms [164].

In this scenario, the consensus has been formulated, that we need to develop "trust" in AI, similar to the trust we have in technologies that have been in use for a longer time. We hence come to the idea of "Trustworthy AI". But what is "Trustworthy AI"? Generally speaking, we can define it as a state of development where we can trust that values and norms orient the design, development and deployment of AI systems, and we potentially have guarantees that international, national, and multinational actors are developing their technologies with respect for fundamental human rights and values [198, 174, 43]. Then, we could genuinely and trustingly integrate AI into society, allowing us to reap the benefits of intelligent automation. For this cause, many areas of intervention, like AI Safety [85], AI Ethics [97], and AI Governance [44], are actively engaged in the development of Trustworthy AI, be that by developing techniques developers can implement in their practice [11, 164], defining and exploring the ethical values that should guide our technological progress [39], or creating the basis for current or future legislations [206].

Principles and values are at the heart of the idea of trustworthy AI development, and we can strive to achieve trustworthiness development if such principles and values are taken into account during the various stages of technological development. Even though ethical principles are many, while also being interpreted in varied ways,[1] there is still agreement about (1) what the most upheld principles in the literature are and (2) what these principles mean in a general sense [41]. For example, we find the principle of privacy at the heart of the EU General Data Protection Regulation, among other data protection regulations worldwide that seek to safeguard individuals' rights and personal data. Meanwhile, reliability and safety support the search for robust AI systems against adversaries and accidents [110]. At the same time, transparency is the normative root of the whole field of explainable AI [146], where we constantly seek ways to provide understandable explanations to black box processes.

While behind every ethical principle, there are human sorrows, issues, or problems, in front of these, we (can) have requirements and tools to implement them practically, when possible. Algorithmic discrimination brings the necessity of fairness [54]. Ecological harm demands the defense of sustainable values [204]. The lack of human autonomy reinforces the need for human control [27]. The absence of reasonable explanations for high-risk scenarios brings forth the requirement for transparency [113]. The invasive nature of our current data-hungry paradigm[2] solidifies pleas for privacy [176].

In this scenario, this work proposes to utilize ethical principles to stipulate minimal ethical requirements for using and developing Trustworthy AI. Drawing from experts in the field, an extensive literature overview, and considering the latest advancements in AI research, we have identified six key areas that, we argue, require critical attention:[3] Fairness, Privacy and Data Protection, Safety and Robustness, Sustainability, Transparency and Explainability, and Truthfulness.

With these values in mind, our work presents to the reader a thorough examination of (1) what these principles are in the context of AI, (2) what requirements can be stipulated so that we can safeguard such values, and (3) how stakeholders can take measures to implement such requirements in their developmental practice. Ultimately, our work can help readers attain the means to aid in developing a healthy and safe AI ecosystem, providing them with a condensed and brief source of ethical and practical guidance for responsible AI development.

---

1 For those interested in exploring the multifaceted landscape of AI ethics, we recommend the Worldwide AI Ethics dashboard by Nicholas Kluge Corrêa et al. [41], which presents an interactive tool for studying the normative discourse in over 200 ethical guidelines and other documents related to AI governance.

2 Scaling laws show the data-hungry tendencies of the foundational model, where the amount of data required to train such artifacts scales as their size increases [88].

3 Values are presented in alphabetical order. This order should not be interpreted as a hierarchy of importance or priority.

# 2 Operationalizable minimum requirements

Many attempts have been made to develop methods to implement ethics in AI development, from developmental frameworks to ethical labels for Trustworthy AI systems [195, 75, 77, 162, 164], all seek to approach the "from principles to practice" problem, i.e., applying ethics in practice, be that in software development or day-to-day business [149, 77]. However, quantifying ethics has some central issues, mainly because easy yes-or-no answers do not exist regarding ethics. Most ethical dilemmas require strong contextualization, while the context might change over time in an ever-changing environment. Meanwhile, moral values are, by nature, abstract concepts that, without a humanistic approach, cannot be adequately translated into practice [50]. Hence, coming up with operationalizable minimum requirements is not trivial and should always be approached through a contextual perspective, i.e., differentially regarding application areas, levels of risk, target population, etc.

Looking at the problem through the lens of the European context, we see that the EU's current approach, as expressed in the AI Act,[4] is risk-based, which means that the degree of regulation that an AI system is subject to depends on its risk class (i.e., minimal, limited, high, and unacceptable), which resonates with other draft proposals [42, 8, 212, 183]. For this work, even though the current mode of governance proposed by the AI Act is differential concerning risk, our propositions are, we argue, general enough to apply to several different risk contexts, with a caveat that for applications and systems classified as "unacceptable risk level", our only suggested requirement is the termination of the system. Meanwhile, to help the reader better frame and assess the possible risk of their system and thus account for the rigor with which our requirements should be fulfilled, we present criteria and examples for every EU AI Act risk level for every principle we work through in this document.

# 3 European Union AI Act: a brief overview

The specific ethical requirements presented in this work are all tailored to be in line with the current EU AI Act.

**However, what is the AI Act?**

The EU AI Act is a regulation proposed by the European Commission on April 21, 2021, and politically agreed upon by the European Commission, the Council of the European Union and the European Parliament on December 8, 2023. This regulation consists of rules for AI system providers and users, which detail each entity's transparency and reporting obligations on the EU market. These requirements apply to European companies and other public and private actors and all AI systems that are being marketed in the EU or that impact people in the EU, regardless of where the systems are developed or deployed. Overall, this legislation is based on ideas of excellence and trust, aiming to promote research and development while guaranteeing safety and the protection of fundamental rights.

Organizations building or using AI systems are responsible for ensuring compliance with the EU AI Act. These compliance obligations depend on the risk an AI system poses to people's fundamental rights, i.e., minimal, limited, high, and unacceptable. Depending on the risk of an application or system, these requirements may involve procedures like:

- Risk assessment surveys to profile the risk of an AI system.
- Compliance assessments that prove a given system is following the AI Act.
- Providing open, transparent, and accessible documentation regarding the AI system.

Currently, most requirements are tied to systems and applications regarded as high-risk. Meanwhile, inability to comply with the rules stipulated by the AI Act may end up generating fines up to €35m for prohibited AI violations (or up to 7% of global annual turnover, whichever is higher), up to €15m for most other violations (or up to 3% of yearly global turnover, whichever is higher), and up to €7.5m for supplying incorrect info Caps (or up to 1.5% of global annual turnover, whichever is higher).[5] The respective lower threshold applies to SMEs and the higher one to other companies for all types of infringements. Hence, adherence to ethical requirements and the AI Act are a reality companies must come to terms with, which, besides avoiding legal fines, may bring market advantages to socially and ethically aware organizations.

---

4 Artificial Intelligence Act.

5 Moreover, other Europe-wide regulations need to be taken into account, such as the General Data Protection Regulation, the Digital Services Act and the Digital Markets Act, which may also result in fines in cases of lack of compliance.

# **4** Compliance and implementation of the suggested assessments

Companies gain numerous advantages by adhering to minimal ethical requirements throughout the life cycle of AI systems. For example:

**1.** Upholding ethical standards enhances brand reputation, fostering trust among consumers and stakeholders. Trust that can translate into increased customer loyalty and market share.

**2.** Ethical AI practices mitigate legal risks, shielding companies from potential lawsuits and regulatory penalties, thereby safeguarding their financial stability.

**3.** Certain requirements, such as the development of procedures to counter model opacity or brittleness, are bound to improve the organization's understanding of the workings and weaknesses of its system while improving overall efficiency and robustness.

**4.** Prioritizing ethics can promote employee morale and retention, attracting top talent to contribute to innovation and growth.

**5.** Trustworthy AI development fosters long-term sustainability by reducing the likelihood of negative societal impacts and ensuring alignment with societal values and expectations, which, in the end, safeguards the interests of stakeholders while cultivating a competitive edge in the rapidly evolving landscape of AI technology.

Hence, such recommendations should be considered advantageous additions to general quality management procedures, compliance mechanisms, or any existing assessment and audit schemes an organization may possess. Moreover, the practices suggested in this work should be considered even for applications where, for the current legislation (EU AI Act), no explicit requirements are established, given that these should not be considered only practices that are valid for compliance reasons but valid for the inherent good and benefits they may bring to the organization and overall society.

· · · · · · · · · · · · · · · · · · · · · · · · · · · · · · · · · · · · · · · · · · ·

Adhering to minimal ethical standards in AI boosts trust,
legal protection, efficiency, employee morale, and
sustainability, driving both compliance and organizational success
while benefiting society.

· · · · · · · · · · · · · · · · · · · · · · · · · · · · · · · · · · · · · · · · · · ·

# **5** Overall Ethical Requirements (O)

As mentioned before, consensus on emerging values and trends in the ethics of artificial intelligence can be found in reviews of the field [222, 97, 81, 41]. While some of these values can be specific to practices that are more easily pin-pointed as requirements (e.g., Truthfulness), others are highly general and resonate with several questions and topics, from philosophy to social sciences and law (e.g., Human-Centeredness, Labor Rights). To better systematize our catalog of requirements hierarchically, we structured four overarching ethical principles, which serve as a foundation for more specific requirements. In short, these serve as foundational pillars to further support recommendations that can shape ethical decision-making processes and ensure that AI systems align with human values, rights, and societal norms:[6]

## Autonomy:

This principle emphasizes preserving human agency and decision-making control in interactions with AI systems. It requires that individuals are adequately informed and empowered to make autonomous choices.

## Beneficence:

The principle of beneficence underscores the ethical obligation to promote the well-being and welfare of individuals and communities through AI technologies. AI development and deployment must prioritize positive societal impacts, such as enhancing human capabilities, fostering fairness and equity, and addressing societal challenges effectively.

## No Harm:

This principle focuses on preventing and mitigating potential risks and harms associated with AI technologies, encompassing physical and societal dimensions. It requires proactive measures to identify, assess, and address risks to individuals, communities, and the environment.

## Accountability:

This principle is a foundational requirement for ethical AI governance and oversight, establishing that we should trace accountability back to ourselves when we violate the above principles, whether through ignorance, incompetence, or ill intent.

In light of these principles, it is essential to consider the following ethical requirements when developing trustworthy AI. In the subsequent sections, we established all other specific requirements of our work as minimal developmental criteria that seek to ground the following overall requirements.

---

6 We highlight that the fulfillment of many of the requirements presented here and in further sections goes beyond the scope of action of developers and engineers, requiring other types of stakeholders, like policymakers, to act in a way that can enable their fulfillment (e.g., by creating policies that will incentivize their realization).

## Requirement O1: Human Oversight and Control

Stakeholders must ensure the integration of human oversight and control mechanisms into AI systems. Hence, they should adopt practices that establish clear human or human-AI intervention procedures, review, moderation, and control, particularly in critical or ethically sensitive situations, to uphold accountability and mitigate risks associated with automated decision-making. Such integration is bound to enhance the transparency and fairness of AI systems and foster public trust and confidence in their responsible use and deployment. Meanwhile, individuals should be informed when they are not interacting with a fellow human being but with a machine, enabling them to decide whether they want to engage with it. AI-generated content must be presented as such.

## Requirement O2: Impact Mitigation and Robust Deployment

Stakeholders must prioritize preventing and remedying AI's negative impacts while ensuring robust deployment of AI systems. They must implement measures to anticipate and address potential adverse effects on individuals, communities, and society. Hence, stakeholders should adopt practices for conducting thorough risk assessments, impact assessments, evaluations, guardrail development, and, when necessary, termination. Furthermore, stakeholders should establish strategies for monitoring and evaluating the performance of AI systems in real-world contexts to facilitate continuous improvement and adaptation. Such proactive measures mitigate risks and contribute to the responsible and sustainable deployment of AI technologies in diverse settings.

## Requirement O3: Open Development and Transparent Reporting

Stakeholders must prioritize open development practices and transparent reporting in AI projects. This requires fostering an environment where collaboration, sharing of insights, and open-source contributions are encouraged and facilitated. Hence, stakeholders should adopt practices to ensure that AI development processes, including data collection, model training, and AI-empowered services, are transparent and accessible to relevant stakeholders. Additionally, stakeholders should establish transparent reporting mechanisms to document AI system capabilities, limitations, and potential biases. Such adherence promotes accountability, facilitates peer review, and fosters public understanding and trust in AI technologies.

## Requirement O4: Privacy, Diversity and Accessibility in, by and for Design

Stakeholders must prioritize integrating privacy, diversity and accessibility considerations into designing and deploying AI systems. This involves implementing practices that incorporate these principles throughout the AI development lifecycle. Privacy measures should ensure stakeholders handle personal data ethically, respecting individuals' rights and expectations regarding their information. Diversity considerations should drive the development of inclusive AI systems representing diverse populations, thus preventing biases and discrimination. Additionally, accessibility by design involves designing AI systems that are usable and accessible to individuals with disabilities, ensuring equitable access to AI technologies. By incorporating these principles from the outset of the conception of a process, product, or service and throughout the research and training phase, operation, and decommissioning, stakeholders can uphold ethical standards, promote fairness, and foster inclusivity in AI technologies, and hence adhere to the principle of "ethics by design".

## Requirement O5: Ethical Stakeholder Inclusion and Co-creation

Stakeholders must include as many vantage points as possible in the design process of AI technologies. This necessitates actively involving all potentially affected stakeholders, such as members of vulnerable groups, in decision-making and development processes. By incorporating diverse perspectives from various stakeholders, including end-users, domain experts, ethicists, and impacted communities, stakeholders can gain comprehensive insights into AI technologies' potential societal impacts and ethical implications. Additionally, fostering an inclusive environment encourages collaboration, increases accountability, and ensures that AI systems meet the needs and values of diverse stakeholders. This inclusive approach ultimately leads to more ethical, equitable, and socially responsible AI deployment.

## Requirement O6: Democracy and Human Rights

Stakeholders should be aware of the extent to which they work with certain institutions. A government, for instance, should verify that a company tasked with services, projects, or the making of AI products complies with human rights and their respective legislation. At the same time, stakeholders should avoid involvement with organizations that contribute to regimes that seek to topple democratic processes or violate human rights. Moreover, stakeholders must actively advocate for incorporating democratic principles and respect for human rights in developing and deploying AI systems. Furthermore, stakeholders should engage in ongoing dialogue and collaboration with civil society organizations, human rights advocates, and democratic institutions to ensure that AI technologies uphold fundamental rights and freedoms.

The development of trustworthy AI necessitates a holistic approach that integrates ethical principles into every stage of design, deployment, and operation. Requirements such as human oversight and control, impact mitigation, transparency, privacy, diversity, stakeholder inclusion, and respect for democracy and human rights serve as guiding pillars for responsible AI development. By adhering to these requirements, stakeholders can uphold accountability, foster inclusivity, mitigate risks, and promote AI technologies' ethical and equitable deployment, ultimately contributing to a more just and socially responsible society.

However, stakeholders must know the inherent trade-offs among such requirements. While the interrelation among the principles that sustain these requirements is undeniable, no rigid hierarchical structure exists between them. Certain historical contexts and specific objectives may occasionally warrant prioritizing one principle or requirement over another, especially in extraordinary circumstances. Consequently, tensions may emerge between these, necessitating thorough discussion and ethical consideration within political and societal discourse. It is imperative to cautiously approach trade-offs between these principles, recognizing that some compromises may be ethically untenable, mainly when fundamental rights and values are at stake.

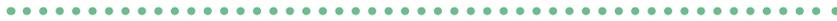

The development of trustworty AI requires a holistic integration of ethical principles such as human oversight, transparency, risk mitigation, diversity, democratic values and human rights to ensure the responsible and equitable deployment of AI technologies.

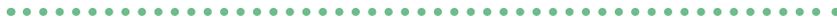

# 6 Fairness (F)

Fairness is an immutable ethical cornerstone, serving as a guiding light across various practical contexts. This principle is rooted in general notions of equity, justice, and impartiality, epitomized by the aspiration for equitable treatment in decision-making and dignity. Hence, we can define fairness as the commitment to ensuring equal opportunities and impartiality for all individuals, irrespective of differences or circumstances, especially on matters outside an individual's control (e.g., where someone was born). In AI ethics, fairness assumes renewed significance as algorithms become the source of unequal treatment and discrimination while increasingly embedded in critical domains of human life. Hence, AI fairness embodies our (i.e., all stakeholders seeking just and fair treatment) concentrated efforts toward rectifying inherent biases entrenched within automated processes driven by AI systems [132, 138].

Within the purview of automated decision-making, there are already many documented cases wherein algorithms exhibit systemic biases, manifesting in outcomes that disproportionately disadvantage certain demographic groups. For instance, in the realm of criminal justice, predictive policing algorithms have been criticized for perpetuating racial profiling and exacerbating existing disparities within the criminal justice system [54]. Moreover, within Generative AI systems, biases can insidiously permeate through the data upon which models are trained, resulting in the generation of content imbued with stereotypes, prejudices, and cultural distortions (e.g., language models generating text that reinforces gender stereotypes or racial prejudices) [109, 197, 129]. These examples underscore the pressing need for ethical vigilance and proactive intervention to rectify algorithmic biases, ensuring that AI systems operate by following societal equity principles.

In AI ethics, concerns regarding fairness are linked to the interplay between sensitive attributes and the behavior of AI systems. Sensitive attributes refer to characteristics such as race, gender, age, socioeconomic status, or any other feature that could potentially lead to discrimination or unequal treatment. Given that sensitive attributes or their proxies[7] are often used as input variables in AI algorithms, how these variables affect the system is a general concern among the AI community. Perhaps it is in issues involving fairness that the old "Garbage in, Garbage out" motto becomes more pronounced [104, 208], something that also points to the fact that algorithmic discrimination is but a reflection of social inequalities which, unfortunately, have no easy technical solution. When algorithms are trained on data that reflects societal biases or historical injustices, they may inadvertently learn and perpetuate those biases, leading to unfair or discriminatory outcomes, especially for specific demographic groups associated with sensitive attributes tied to those unfair cases.

Now, it is imperative to clarify that the scope of this work does not aim to prescribe solutions or establish requirements for addressing social inequalities [172, 182, 155, 66]. Instead, our primary objective lies in fostering strategies geared toward detecting, preventing, and mitigating algorithmic discrimination. Hence, our ethical requirements should be seen as something other than a catch-all solution to such a complicated and systemic problem, which, in the end, is a matter of concern that pre-dates the era of algorithms and artificial intelligence.

To help organizations prospect the level of risk regarding their system or application, we propose the following criteria, in line with the four main categories of risk identified in the EU AI Act (examples in Table 1):

1. **Minimal:**

   Systems with minimal risk regarding the principle of fairness do not process sensitive information or their proxies in its workings.

2. **Limited:**

   Systems with a limited risk regarding the principle of fairness process sensitive information or their proxies in its workings but are not used for applications tied to critical services (e.g., healthcare, policing, finance).

3. **High:**

   Systems that present a high risk regarding the principle of fairness process sensitive information or their proxies in its workings and are used for applications tied to critical services.

4. **Unacceptable:**

   Systems that present unacceptable risk regarding the principle of fairness process sensitive information or their proxies in its workings and are used for applications that can violate fundamental human rights and values.

---

7 Proxies refer to indirect or surrogate variables used in place of sensitive attributes that may lead to discrimination outcomes in algorithmic decision-making. For example, a name can be a proxy for gender, while an address can be a proxy for race in racially segregated communities.

Table 1: Examples of application areas and their categories of risk (Fairness)

| | |
|---|---|
| **Minimal:** | Weather Forecasting<br>AI Art Generation<br>Traffic Management |
| **Limited:** | Music Recommendation<br>Advertising Algorithms<br>Online Retail |
| **High:** | Hiring Support<br>Financial Fraud Detecetion<br>Traffic Management |
| **Unacceptable:** | Facial Recognition for Public Surveillance<br>Advertising Algorithms<br>Online Retail |

The following minimal ethical requirements tie normative recommendations to implementable practices. All requirements are general and should be employed regardless of the risk category of a system.

## Requirement F1: Bias Analysis

Performing a bias analysis is the foundational step in improving the fairness of an AI system. By systematically examining data, features, and decision-making processes, a bias analysis helps uncover patterns of discrimination or unfairness that may be encoded within the algorithm. This initial assessment provides crucial insights into the sources and manifestations of bias, enabling stakeholders to formulate targeted strategies for mitigation and remediation. Moreover, bias analysis fosters transparency and accountability, allowing for informed decision-making regarding deploying and using AI systems.

To infer if an AI system possesses a discriminatory bias against a particular group, we can take two main approaches: statistical or causal. The statistical approach involves analyzing patterns and correlations within the data to identify disparities in outcomes across different demographic groups.[8] We can also extend this approach to AI systems by sampling comparisons of their input-output behavior in different groups of interest. For example, using fairness metrics or heuristics[9] to determine if a system is biased toward a particular outcome is an already established practice in machine learning Fairness, giving ready-to-use metrics that can inform stakeholders on the "fairness status" of a system [63, 207, 35, 30, 138, 13].[10] For example, if a credit scoring system provides good scores to an ethical majority 85% of the time but only 50% the time to an ethical minority, this might indicate a violation of statistical parity and, hence, a fairness violation.

While these methods may assess the existence and magnitude of bias, they may fail to establish a causal connection to individual instances. In contrast, the causal approach aims to uncover the causes driving biased outcomes by examining the causal relationships between variables and outputs [217, 133, 28, 153]. By employing techniques such as causal inference or experimentation, this approach enables a deeper understanding of how certain factors contribute to disparities and allows for developing targeted interventions to address the root causes of bias. For example, suppose a text-to-image model outputs an image of a man working at a hospital when given the prompt "A working man" but produces an image of a waiter when we modify the prompt to "A working black man". Through a causal analysis, in that case, we can diagnose a discriminatory bias in such a system.[11]

Systematic scrutiny of data, features, decision-making processes, and general system behavior can reveal biases, empowering stakeholders to devise targeted strategies for mitigation. Whether employing statistical methods to detect disparities or causal approaches to unveil underlying causes, bias analysis fosters transparency and accountability in AI deployment.

## Requirement F2: Open Disclosure

In tandem with bias analysis, open disclosure is a pivotal requirement in ensuring that the inherent biases of AI technologies are made clear to a broad audience of stakeholders. Hence, openly disclosing possible biases should be considered a standard practice and an absolute ethical requirement for high-risk applications. This transparency not only cultivates trust but also empowers stakeholders to assess the fairness and reliability of AI systems. Furthermore, open disclosure encourages collaboration and accountability, enabling stakeholders to collectively address and mitigate potential biases and ethical concerns. As documentation is already part of the work of organizations producing state-ofthe-art AI systems, AI applications need to be accompanied by documentation that discloses information about the possible biases systems can have [181, 163].[12]

## Requirement F3: Rectify or Mitigate Biased Outcomes

While bias analysis and open disclosure are initial steps crucial for identifying potential biases, they're just the beginning. It's equally vital to proactively address these biases, especially given the high stakes involved for equality and fairness. In a city where historically marginalized neighborhoods are underserved by public transportation, for instance, an AI-powered traffic management system trained with biased samples could inadvertently exacerbate existing disparities if not carefully monitored and corrected. So, we must actively intervene beyond recognizing biases, emphasizing this requirement's need. In short, rectifying or mitigating biased outcomes entails implementing interventions and strategies to minimize the disparate impact on marginalized or underrepresented groups. To achieve this, we can focus on two main areas of interference (regarding this requirement): data and system.

---

8 Tools like Dalex, AI Fairness 360, Fairlearn, REVISE, and What-If can help you automate several processes related to bias analysis, like dataset exploration.

9 Disparate impact, or the "80% rule", is a heuristic (and statistic) form to evaluate discrimination. To learn more, we recommend "Certifying and removing disparate impact".

10 To learn more on how to apply different fairness metrics, we recommend "Fairness definitions explained".

11 To learn more about causal fairness and how to apply it, we recommend "Causal Conceptions of Fairness and their Consequences", which is also accompanied by practical demonstrations (Available on GitHub).

12 Examples of AI systems that reveal the presence of potential biases can be found here and here. As a reminder, it is essential to note that filling out a model card requires input from different roles, like the developer (e.g., a technician who runs and writes the code), the socio-technic (e.g., someone skilled at analyzing the interaction of technology and society long-term), and the project organizer (e.g., one who understands the overall scope and reach of the technology). For more details on how to structure and build such documentation, we recommend the "Annotated Model Card Template".

Data quality is paramount when developing an AI system through a learning paradigm. Quality and representativeness significantly influence AI systems' fairness and equity, as already shown by studies that strived to create fair datasets for sensitive areas of application [102].[13] Therefore, meticulous examination and management of training data are essential. Techniques such as data augmentation [185, 158], sampling methodologies [16, 117], and dataset balancing [218, 82] can be instrumental in ensuring a more equitable representation of diverse demographic groups within the training data.[14] Moreover, ongoing monitoring and validation of data inputs are crucial to detect and address emerging biases over time.

In addition to data management, interacting directly with a model and implementing guardrails is vital for rectifying or mitigating biased outcomes in AI systems. In this context, model guardrails encompass techniques and strategies embedded within the system to monitor and minimize biases during inference stages. These guardrails serve as safeguards against perpetuating or amplifying biases in the data. Techniques such as adversarial debiasing [223], fairness-aware regularization [101], and fairness constraints [220] can be integrated into a model or its larger system to promote fairness and equity.[15] By incorporating model guardrails into AI systems, stakeholders can proactively address biases, uphold ethical standards, and promote fairness and equity across various applications and domains. By embracing these approaches, stakeholders can navigate the complexities of AI development with a commitment to fairness and equity.

## Requirement F4: Human Moderation

Besides all the previously established requirements, specifically in high-risk applications, the path toward automation must be accompanied by scalable human oversight and moderation [15, 29, 112]. While automated techniques and algorithms are crucial in identifying and mitigating biases, in high-risk scenarios (e.g., recidivism prediction instruments), human intervention is essential in ensuring ethical and responsible AI deployment, given the abstract, nuanced, and ever-changing landscape of human normativity, which constantly clashes with AI systems as out-of-distribution cases. In short, human moderation provides a critical layer of scrutiny and accountability, allowing for the nuanced evaluation of complex ethical considerations and contextual nuances that automated systems may overlook. Moreover, human moderators can offer insights into the socio-cultural implications of AI technologies and contribute to developing contextually sensitive solutions that align with societal values and norms. For instance, consider using automated recidivism prediction algorithms in the criminal justice system. While these algorithms may excel in processing large datasets and identifying patterns, they need support to account for the complexities of individual cases and the societal context in which they operate. A human moderator, familiar with the nuances of the legal system and sensitive to social dynamics, could recognize when the algorithm's recommendations might inadvertently perpetuate biases or unfairly target specific demographics.

For minimal or limited risk applications, automated mitigation strategies that use human-AI collaboration strategies can help scale human oversight to realms where human moderation does not scale (e.g., moderation of online forums) [31].[16]

By integrating moderation into the AI development lifecycle, stakeholders can foster greater accountability in the use of AI systems, ultimately advancing the goal of creating fair, inclusive, and socially responsible AI technologies.

---

13 For example, FairFace is a face image dataset that is race-balanced. It contains 108,501 images from 7 different race groups.

14 Tools like AI Fairness 360 and REPAIR have in-built methods to repair datasets to more equal distributions.

15 One can also include automated moderation techniques to block certain types of behavior of, for example, generative systems, e.g., specifying lists of banned terms, NSFW detection tools, and overall moderation APIs.

16 To learn more about AI and human-AI moderation, we recommend "Human-AI Collaboration via Conditional Delegation: A Case Study of Content Moderation".

# 7 Privacy and Data Protection (P)

Privacy is a cornerstone value, often endangered when personal data undergoes processing. When we refer to personal data, we speak of information wielded to pinpoint an individual's identity. However, privacy surpasses mere data; it extends into realms such as decision-making and human autonomy. Safeguarding one's privacy nurtures independence, fostering the capacity to manifest their true self. It carves out a sanctuary wherein individuals can thrive, expressing themselves freely. Although privacy concepts are ubiquitous across cultures, liberal societies prioritize privacy for fostering individual development. Consequently, it assumes a pivotal role in the framework of democratic states, comprising citizens with diverse perspectives. In essence, safeguarding privacy intertwines closely with (cyber)security measures. The mechanisms that shield a system are akin to those that fortify against data breaches or loss.

While data protection legislation is prevalent worldwide, the emergence of artificial intelligence systems introduces novel inquiries owing to the vast quantities of data they assimilate for training and operation. Moreover, the adeptness of artificial intelligence systems in establishing connections among previously disparate data sources [43] or extracting profound insights from processed data often surpasses human capabilities [93]. Consequently, distinct challenges manifest concerning data processing by autonomous systems. Furthermore, data acquisition through sensors integrated into AI-driven products exacerbates the opacity surrounding data collection.

The scope of data has evolved beyond mere alphanumeric characters stored in text files; it now encompasses diverse formats such as voice recordings, images, videos, and similar media, potentially containing personal information. Processed data may not exclusively comprise personally identifiable information; it could also include data about system performance, which warrants protection. For instance, performance data might divulge insights into a system, encroaching upon operational secrets [43]. Risks arising from the data processing by artificial intelligence systems and their training transcend concerns solely about individual natural persons. They extend to groups, organizations, institutions, and enterprises, with the latter facing potential harm in scenarios like data breaches jeopardizing intellectual property. Moreover, deploying technologies such as facial recognition in public spaces can infringe upon an individual's privacy.

For global enterprises seeking to engage in various markets worldwide, navigating regional legislation becomes imperative. In Europe, for instance, adherence to EU law, specifically the General Data Protection Regulation, is requisite when dealing with data subjects within EU territory [1, 87]. However, exemptions exist regarding data protection for research purposes. Synthetic data emerges as a viable alternative, circumventing certain challenges associated with real-world data usage. Companies may thus contemplate employing artificial data, referring to (annotated) information that mirrors real-world scenarios and is generated, for instance, through computer simula-tions.[17]

Certification in data protection, exemplified by seals and marks,[18] serves as tangible evidence of compliance with data protection principles, as in accordance with article 42 of the EU's GDPR [1]. Documenting compliance with data protection standards is essential and meticulous [87]. The GDPR explicitly prohibits the processing of special categories of personal data, encompassing information revealing sensitive aspects such as racial or ethnic origin, political opinions, religious or philosophical beliefs, trade union membership, genetic data, biometric data for unique identification, data concerning health, or data regarding a person's sexual life or orientation, unless exceptions outlined in Article 9(2), such as data subject consent, are applicable [1].[19]

Beyond safeguarding personal data, ensuring the quality and integrity of datasets is imperative [4], necessitating organizational investment in data governance. Instances like the notorious case of an AI system mislabeling black individuals as gorillas underscore the importance of rectifying flawed datasets. However, from a privacy-preserving perspective, the resolution cannot entail continually augmenting system training data or incorporating data from vulnerable populations. Such an approach would perpetuate the accumulation of extensive datasets, including sensitive information from vulnerable demographic groups.

Hence, as we navigate the landscape of privacy and data protection in the age of artificial intelligence, it becomes evident that safeguarding individual autonomy and upholding democratic principles remain paramount. The evolution of data formats and the proliferation of AI systems pose novel challenges, necessitating robust frameworks and innovative solutions. Yet, amidst these complexities, the imperative to respect privacy transcends mere compliance. It embodies a commitment to fostering trust, empowering individuals, and preserving fundamental human rights in the digital era.

Privacy is crucial for autonomy and democracy, facing new risks from AI's vast data processing, requiring strong protections to maintain trust and rights.

---

17 For example, MIT's Data to AI Lab has developed a library for synthetic data generation for tabular data, now maintained by a company named DataCebo.
18 Some examples of such seals are TrustArc (formerly TrustE) and EuroPriSe.
19 Similar distinctions are also present in other legislations, such as the Brazilian GDPR [57].

To help organizations prospect the level of risk regarding their system or application, we propose the following criteria, in line with the four main categories of risk identified in the EU AI Act (examples in Table 2):

**1. Minimal:**
Systems with minimal risk regarding the principle of privacy do not need to process personally identifiable information for their basic functioning in any way.

**2. Limited:**
Systems with a limited risk regarding the principle of privacy process personally identifiable information in their workings but are not used for applications tied to critical services (e.g., healthcare, policing, finance).

**3. High:**
Systems that present a high risk regarding the principle of privacy process personally identifiable information in their workings and are used for applications tied to critical services.

**4. Unacceptable:**
Systems that present unacceptable risk regarding the principle of privacy process personally identifiable information in their workings and are used for applications that can violate fundamental human rights and values.

Table 2: Examples of application areas and their categories of risk (Privacy)

| Minimal: | Weather Forecasting<br>Industrial Automation<br>Translation Engines |
|---|---|
| Limited: | AI Assistant<br>Email Filtering<br>Recommendation Engine |
| High: | Healthcare App<br>Fraud Detection<br>Targeted Marketing |
| Unacceptable: | Social Credit Scoring<br>Autonomous Weapons<br>Facial Recognition for Public Surveillance |

---

20 For example, MIT's Data to AI Lab has developed a library for synthetic data generation for tabular data, now maintained by a company named DataCebo.

The following minimal ethical requirements tie normative recommendations to implementable practices.
All requirements are general and should be employed regardless of the risk category of a system.

## Requirement P1: Purpose Limitation and Data Minimization

Developers and implementers of artificial intelligence systems must question the necessity of processing personal data for system functionality. They should consider whether training or operating the system is feasible without relying on personal data. This imperative aligns with the principle of data minimization, which dictates that data processing should always be appropriate, pertinent, and constrained concerning its processing objective [1, Art. 5 (1) (b,c)]. Data minimization and purpose limitation should be coupled with a reflective examination of the true objective of the AI software [87].

## Requirement P2: User Control and Information

Individuals must receive notification when their data undergoes processing. Users should be able to control their data to the greatest extent possible. Meaningful and informed consent is paramount; pre-checked checkboxes do not constitute valid consent [87]. Users should have the option to opt in or, at the very least, opt out of data processing. Moreover, following the GDPR, users retain the right to revoke their consent to data processing at any juncture [1, Art. 7 (3)]. The GDPR further establishes the "right to be forgotten," entitling individuals to request the deletion of their data, as stipulated in Article 17 [1, Art. 17].

Understanding the perspectives of customers or data subjects is invaluable for specific stakeholders, enabling alignment between expectations and the actual technological capabilities of a system [93]. Users should also be kept abreast of governmental and law enforcement involvements concerning an AI system. Information dissemination should prioritize accessibility, employing user-friendly formats like pictograms or privacy dashboards [87]. A comprehensive privacy policy should cater to external users and internal stakeholders such as employees and company members.

## Requirement P3: Data Protection by Design and Default

Data protection must be ingrained from the outset in the development of AI systems, adhering to the principle of implementing privacy protection by design. This entails incorporating various technical and organizational measures during the system's conceptualization phase. These measures may encompass anonymization or pseudonymization (cf. requirement P4) and differential privacy—a concept aimed at ensuring privacy for individuals' information within a group database.[21]

Additionally, data protection should be automatically guaranteed as the default setting. This entails preconfiguring systems with the most privacy-friendly options, safeguarding users' privacy, and preventing inadvertent disclosure of their data to unknown entities without consent. Data protection by design and by default constitutes an obligation for data controllers as stipulated by the GDPR [1, Art. 25].

## Requirement P4: Anonymization and Pseudonymization

To prevent data breaches, anonymization[22] or, at the very least, pseudonymization[23] techniques can be used when processing personal data. When information is anonymized, data protection regulations cease to apply [1, Recital 26]. However, it's essential to implement technical measures to ensure that re-identification remains unfeasible, as technological advancements or additional information releases could reverse anonymization. Preventing unwarranted and unforeseen cross-correlation of datasets is crucial [93]. Anonymization can also serve the data controller's interests, particularly if a user later withdraws their consent to data processing. This approach can avert the costly retraining of the entire model in a worst-case scenario [164]. All data, including metadata that may facilitate the identification of an individual [87], should be promptly deleted once they are no longer required. This principle extends to scenarios where the system is decommissioned [93].

## Requirement P5: Data Protection Impact Assessment

A Data Protection Impact Assessment (DPIA) should be conducted to evaluate potential data protection risks inherent in a project, particularly when incorporating new technologies [1, Art. 35, Recital 90]. Various tools and guides are available to facilitate the execution of a DPIA [136, 37], emphasizing its necessity for sensitive data processing scenarios like profiling, public space monitoring, or automated decision-making. Stakeholders, including developers and customers of AI systems, can refer to a list compiled by the European Data Protection Supervisor to determine the need for a DPIA [56]. All relevant parties must prioritize the protection of potentially affected individuals' privacy. Simply relying on the AI system provider to safeguard users' data is insufficient, as harm may arise upon system deployment [55]. Consequently, developers/sellers and buyers of AI systems should conduct DPIAs. Additionally, prospective positive outcomes from developing and deploying AI systems should be considered [202]

---

21 Different tools from different actors, including Google and IBM, are available for using differential privacy as a privacy-preserving technique, e.g., OpenDP.

22 Anonymization is the process of removing any information that can link sensitive personal data back to an individual entity. Learn more in "Anonymity Set".

23 Psuedonymisation is a process of removing or replacing personal identifiers from data and using placeholder values or reference numbers instead. Learn more in "Psuedonymous Identity".

## Requirement P6: Codes of Conduct and Privacy Management

Developers and implementers of AI systems must engage in introspection regarding their business practices. At the same time, data processors should formulate codes of conduct [1, Art. 40]. At the management level, individuals should routinely contemplate the company's values, fostering inclusivity by involving staff members. Privacy assurance should be multifaceted, incorporating both technical and organizational measures. A well-defined privacy policy, structured as a twofold approach encompassing an "enforce strategy" aimed internally at the organization and ensuring compliance with externally communicated privacy statements, is essential [87]. Collectively, these components constitute a comprehensive "privacy management system" [87], which can be implemented in tandem with a DPIA. This system should also encompass potential mitigation measures for addressing risks and breaches [93]. Furthermore, a broader data management strategy can aid in anticipating undesirable outcomes and contribute to ensuring the quality of datasets.

# 8 Safety and Robustness (SR)

The concepts of safety and robustness are closely related to another ethical principle: non-maleficence [41], which is a principle commonly used in the context of bioethics and medical ethics [145, 40, 72], dictating that medical practitioners and bio researchers must do no harm or allow harm to be caused to a patient through neglect. Similarly, safety and robustness are considered an indispensable pillar of AI ethics [7, 100, 85, 89], dictating that AI systems should have guarantees in relationship to their safety (avoidance of harm) and robustness (remain safe and accurate under adverse conditions). Such requirements are paramount when high-stakes applications meet the complex, dynamic, fuzzy, and noisy real world.

In terms of safety, what we are mainly concerned with is avoiding unintended harm, i.e., accidents.[24] In almost any engineering field, we can define accidents as situations where a human designer has a specific objective or task in mind. Still, the system designed and deployed for that task produced harmful and unexpected results [7]. In the context of AI, these events usually come in two primary forms:

- Side Effects: unintended adverse effects that the system designers did not originally envision (e.g., a recommendation system that produces addictive behavior).
- Vulnerabilities: harmful behaviors that adversaries can elicit to produce negative consequences (e.g., when someone hacks into an AI system to perform illicit activities).

Unlike side effects, vulnerabilities may have an intentional component tied to the desires of the adversary. Still, they can also happen due to distributional shifts, where vulnerabilities or hidden functionalities emerge when a system operates outside its original usage scope. That is when the concept of robustness becomes imperative. For AI systems to be considered safe, they must also be robust under adversarial or unusual scenarios. Hence, both these principles are united under this set of specific requirements. Unfortunately, as is the case in other safety-critical areas [193, 99, 64], robust safety regulations and practices are almost only considered after a tragedy.[25] Our efforts in listing these minimal ethical requirements, in line with the AI Act, are to help organizations think in a preventive matter, which is bound to improve the workings of their technologies and the safety of all stakeholders involved.

To help organizations prospect the level of risk regarding their system or application, we propose the following criteria, in line with the four main categories of risk identified in the EU AI Act (examples in Table 3):

**1. Minimal:**

Systems that present a minimal risk regarding the principle of Safety and Robustness are those where failure or malfunctioning of the system would not pose any direct threat to human safety or cause significant disruptions. These systems may operate in non-critical environments or have fail-safe mechanisms that mitigate potential risks to an acceptable level.

**2. Limited:**

Systems that present a limited risk regarding the principle of Safety and Robustness are those where failure or malfunctioning could potentially lead to minor injuries or inconveniences to individuals. While these systems are not directly tied to critical safety concerns, their failure could still adversely affect users or stakeholders.

**3. High:**

Systems that present a high risk regarding the principle of Safety and Robustness are those where failure or malfunctioning could result in significant harm, injury, or loss of life. These systems are often found in critical infrastructure, transportation, healthcare, or manufacturing sectors.

**4. Unacceptable:**

Systems that present an unacceptable risk regarding the principle of Safety and Robustness are those where failure or malfunctioning could lead to catastrophic consequences, including widespread loss of life, severe environmental damage, or irreversible harm. These systems should not be developed or deployed under any circumstances, and existing deployments should be terminated immediately.

---

24 In this section, we are not considering cases where malicious actors intentionally use AI systems to provoke harm, given that such applications require only one ethical requirement (prohibition and termination). For readers interested in the landscape of potential security threats from malicious uses of artificial intelligence technologies, we recommend "The Malicious Use of Artificial Intelligence: Forecasting, Prevention, and Mitigation".

25 As the saying goes for the Aviation industries: "Aviation regulations are written in blood".

Table 3: Examples of application areas and their categories of risk (Safety and Robustness)

| | |
|---|---|
| **Minimal:** | Auto Correct<br>Translation Engines<br>Optical Character Recognition |
| **Limited:** | Text-to-Image Models<br>Assisted Image Editing<br>Recommendation Systems |
| **High:** | Weather Forecast<br>Assisted Policing<br>Assisted Surgery |
| **Unacceptable:** | Automated Policing<br>Recidivism Forecasting<br>Automated Cyberattacks |

The following minimal ethical requirements tie normative recommendations to implementable practices.

All requirements are general and should be employed regardless of the risk category of a system.

## Requirement SR1: Robust and Open Evaluation

The first step towards safe and robust AI is to thoroughly evaluate our systems. As the potential risks associated with AI applications increase, so should the rigor of evaluation efforts. The evaluation of AI systems concerning safety and robustness must be conducted with meticulous attention to detail and openness to scrutiny. Hence, systems categorized as minimal risk may afford more flexibility in evaluation methodologies, and applications operating in high-risk domains necessitate a more comprehensive and transparent assessment process.

For example, in the case of machine learning systems, while the vanilla train/validation/test split may be sufficient to access the reliability of an innocuous system that will operate under a very narrow environment, it may fail to explore the limitations of systems that have to work in more lose and unpredictable environments. Hence, a minimal requirement for any system categorized as high-risk should be evaluating its functionality on standard benchmarks. For organizations that do not possess their own harness of evaluations, open evaluation benchmarks should be utilized (e.g., Language Model Evaluation Harness for text generation models [69], FACET for facial recognition applications [80], Q-Bench for multimodal model [215]). However, robust and reproducible evaluation should be a standard practice, regardless of the level of risk. Hence, organizations are encouraged to adopt standardized evaluation frameworks and methodologies tailored to the specific characteristics of the AI system under scrutiny. Additionally, leveraging community resources, such as benchmark datasets, evaluation protocols, and collaborative platforms, helps to stimulate a culture of openness and collaboration in the AI community.

Finally, high-risk systems should be incentivized to make evaluation procedures and results accessible to relevant stakeholders, including developers, regulators, and end-users. Exposing benchmark results is already standard practice in academia and is being adopted by the industry.[26] Moreover, minimal performance on such evaluations may become required for high-risk cases as regulation becomes more rigorous. It is imperative to prioritize robust and open evaluation practices, particularly in high-risk applications such as autonomous vehicles, medical diagnosis systems, and critical infrastructure control systems. By adhering to rigorous evaluation standards and embracing transparency, stakeholders can instill confidence in the safety and reliability of AI technologies, thereby promoting their responsible development and deployment.

## Requirement SR2: Red Teaming and Broader Impact Analysis

Organizations that develop or use AI systems may be unaware of their systems' vulnerabilities or hidden functions when dealing with (1) the complexities of the real world or (2) the ingenuity of adversaries. Hence, ensuring the safety and robustness of AI systems requires proactive measures to identify these vulnerabilities, assess potential risks, and mitigate adverse impacts. That is where the idea of employing red teams comes into place.

The term red team, as the name suggests, comes from the United States of America and Cold War-era military simulations (adversarial teams were the "red team") [124]. Generally, it means role-playing as the adversary while conducting a vulnerability assessment. In other words, good guys pretend they are bad guys to the best of their ability, so when the actual bad guys come, the good guys are prepared. Nowadays, red teaming is a common practice in cybersecurity, where "red-teamers", also referred to as pen(etration) testers, are hired to test the security of physical locations or computer networks [2]. Finally, the AI community has adopted this practice, given that modern AI systems based on large neural networks may exhibit emergent properties and behaviors that designers cannot fully predict ahead of time.

In the context of AI, red teaming involves systematically simulating adversarial attacks, scenarios, and misuse cases to uncover weaknesses in AI systems' behavioral processes [68]. By adopting the perspective of malicious actors or unintended users, red teams can identify vulnerabilities that traditional testing methodologies may overlook. These exercises encompass various threat vectors, including adversarial manipulation [74], data poisoning [189], back door attacks [91], and other general misbehaviors (e.g., unintended biases) and patterns that can be uncovered through adversarial techniques [110]. It is essential to notice that red teaming practices are not limited to machine learning systems, given that other types of intelligent systems can also be corrupted if the larger context in which they operate is insecure. Also, given that machine learning systems do not live in a vacuum, questions related to general software and system security should also be areas to explore in the red teaming process.

Red teaming should be conducted iteratively throughout the development life cycle to validate the effectiveness of mitigation strategies and resilience measures. It's crucial to acknowledge that red teaming is a process that varies for each system, depending on its complexity, domain, and potential threat landscape. Hence, organizations should carefully adopt measures sensitive to their context, ensuring that red-teaming efforts are tailored to the risks and challenges faced by the AI system under scrutiny.[27] For applications categorized as minimal risk, automated forms of red teaming, such as adversarial training [214, 184, 9], may suffice to uncover and address vulnerabilities. However, in the case of high-risk applications, human-led red teaming exercises become indispensable (in addition to automated forms [161]), providing a more nuanced understanding of potential threats and ensuring comprehensive risk assessment.

Finally, reporting the findings of a red teaming process should be a requirement for high-risk situations. Currently, model reporting practices also integrate the disclosure of systems limitations and potential flaws, much like pharmaceutical drugs have warnings against substance abuse. These broader impact reports should disclose the potential social, ethical, and environmental ramifications of deploying a given AI system in real-world settings. Beyond technical metrics, organizations must consider the broader implications of their AI solutions on stakeholders, communities, and society at large (e.g., impact expectations on the job market due to technological displacement). This analysis should encompass fairness, accountability, transparency, privacy, and inclusivity to ensure that AI deployments align with ethical principles and societal values.[28]

---

26 The existence of leaderboards is an example of this (e.g., HELM and DecodingTrust leaderboards.).
27 The study "Red Teaming Language Models to Reduce Harms: Methods, Scaling Behaviors, and Lessons Learned" offers a comprehensive review of the challenges related to building and administrating red teams.
28 Here, we have two examples, GPT-3 and Llama, of model cards that disclose limitations and ethical considerations.

## Requirement SR3: Human Oversight

As AI applications become increasingly pervasive and complex, human intervention and decision-making are critical in mitigating risks, addressing uncertainties, and upholding ethical standards. According to the EU High-Level Expert Group on AI (EU HLEG) [4], any "allocation of functions between humans and AI systems should follow human-centric design principles and leave meaningful opportunity for human choice", which entails implementing human oversight and controls over AI systems and processes. Meanwhile, the EU AI Act states that AI should "be designed and developed in such a way, including with appropriate human-machine interface tools, that they can be effectively overseen by natural persons during the period in which the AI system is in use". Therefore, human oversight becomes a requirement to ensure AI systems' safety, robustness, and moral integrity are upheld throughout their lifecycle, tying a human factor to their existence.

While AI systems with minimal risk may have more lenience regarding oversight, deploying AI systems in highstakes domains necessitates human oversight to complement automated processes and algorithms. Human oversight encompasses a range of activities, including monitoring system behavior, interpreting outputs, making critical decisions, and intervening when necessary to prevent or mitigate adverse outcomes. In short, human moderators can provide contextual understanding and domain expertise while supporting the identification of edge cases, ambiguous scenarios, and ethical dilemmas that may challenge the robustness and fairness of AI systems.

According to the EU HLEG, the following levels of oversight should be considered:

- Human in the loop: human intervention at every stage of the AI lifecycle.
  For example, a radiologist may review and interpret the AI-generated diagnostic in a medical diagnosis assisted by an AI system, providing additional context and expertise before finalizing the diagnosis and treatment plan.

- Human on the loop: human intervention during the design cycle of the system and monitoring the system's operation. For example, in an automated content moderation system for a social media platform, human moderators periodically review flagged content to ensure accuracy and fairness while providing feedback to improve the algorithm's performance.

- Human in command: the capability to oversee the overall activity of the AI system and decide when and how to use the system in any particular situation. For example, in an autonomous vehicle, the human driver retains ultimate authority over the vehicle's operation, with the ability to override the AI system's decisions in emergencies, such as taking manual control to avoid a collision.

Organizations should establish clear lines of responsibility, accountability, and authority for human decision-makers to operationalize these approaches by developing training programs, guidelines, and protocols to empower human stakeholders with the knowledge, skills, and resources needed to fulfill their oversight roles effectively and at scale.

# 9 Sustainability (SU)

The relentless march of technology within modern society inexorably intertwines with ecological markers, profoundly influencing our planetary boundaries regarding what our world can support [59, 157, 178, 94, 213]. Technological advancements, while promising efficiency and convenience, often come at a steep cost to the environment. From the extraction of rare earth metals to the proliferation of electronic waste, the ecological footprint of our digital age is unmistakable. At the same time, given that the current advances of AI are tied to the severe scaling of its components, like the amount of hardware (and resources to run this hardware) available, many currently question the sustainability of AI development at the scale it is being pursued [154, 21, 177, 62, 204].

If we narrow our focus to artificial intelligence, it becomes apparent that the most significant ecological impacts stem from developments tied to the deep learning paradigm [114, 73, 166]. Training massive neural networks necessitates substantial computational power [125, 111, 159, 52], driving the demand for specialized hardware optimized for such tasks. However, this quest for efficiency often comes at the expense of environmental sustainability. The carbon footprint of training large neural networks is substantial, with estimates suggesting that training a single deep learning model can emit as much carbon dioxide as several cars over their lifetimes [191, 128]. Additionally, the operation of data centers to support these computational tasks requires vast amounts of water for cooling, further straining already stressed water resources [150].

Moreover, the production of specialized hardware relies heavily on extractivist practices, contributing to environmental degradation and social injustices in mining communities [83, 152, 148]. For instance, cobalt mining, a crucial component in many computing devices, has been linked to child labor and environmental pollution in regions like the Democratic Republic of Congo [201, 165, 12]. These interconnected environmental costs underscore the urgent need for reevaluation and mitigation strategies in pursuing AI development, challenging the prevailing narrative of technological progress at any cost.

These pressing environmental concerns must be carefully considered when developing trustworthy AI. As stakeholders in the AI field strive to ensure the ethical and responsible deployment of artificial intelligence systems, environmental sustainability must be integrated into the core principles of AI development. Failure to address these ecological impacts exacerbates environmental degradation and poses significant risks for (not so) future generations. Without concerted efforts to mitigate the ecological costs of AI development and the technological industry in general (not to mention the clothing [36], agricultural [190], and mining industries [141]), we risk leaving behind a legacy of depleted resources, polluted environments, and social injustices for future inhabitants of our planet. Thus, current stakeholders must acknowledge and address these challenges, paving the way for a more sustainable and equitable future where technological progress is harmonized with ecological well-being.

Now, to promote sustainable AI development, there are two main approaches stakeholders can take [204]:[29]

> • AI for sustainability: applying artificial intelligence technologies to address sustainability challenges and promote sustainable development.[30]
> • Sustainability of AI: developing and deploying artificial intelligence in an environmentally responsible manner.

This work focuses on the second approach. Hence, all our suggested requirements are made with the sustainability of AI in mind. To help organizations prospect the level of risk regarding their system or application, we propose the following criteria, in line with the four main categories of risk identified in the EU AI Act (examples in Table 4):

---

29 If you wish to explore the requirements for this distinction further, we recommend "Challenging AI for Sustainability: what ought it mean?".
30 To learn more about research avenues for using AI to achieve sustainable outcomes, we recommend "Artificial intelligence for sustainability: Challenges, opportunities, and a research agenda".

**1. Minimal:**

Systems that present a minimal risk regarding the principle of
Sustainability require a negligible amount of energy to be produced and used.
For these, the threshold is set to be between 1 - 50 KgCO2eq during development
(i.e., around 1000 Km driven by an average gasoline-powered passenger vehicle.) and
below 50 g CO2 per usage (i.e., an input-output mapping of the model).[31]

**2. Limited:**

Systems that present a limited risk regarding the principle of
Sustainability remain within a manageable level of energy consumption and environmental impact.
This threshold is set to be between 50 - 250 KgCO2eq during development (i.e., up to 4500 Km driven
by an average gasoline-powered passenger vehicle.) and below 100 g CO2 per usage.

**3. High:**

Systems that present a high risk regarding the principle of Sustainability require specialized
hardware accelerators for development or use and produce a non-negligible environmental impact.
This threshold is set between 250 and 50,000 KgCO2eq during development (i.e., inside the range
for the documented emissions of large neural networks)[31] and below 200 g CO2 per usage
(i.e., twice the allowance given to the limited category).[32]

**4. Unacceptable:**

Systems that present an unacceptable risk regarding the principle of Sustainability could generate
emissions above 50 tons of CO2eq in a single month of training or use.

Table 4: Examples of application areas and their categories of risk (Sustainability)

| | |
|---|---|
| **Minimal:** | BERT<br>ResNet |
| **Limited:** | GPT2<br>ViT-Large |
| **High:** | BLOOM<br>SDXL-Turbo |
| **Unacceptable:** | > 50 tCO-2eq<br>> 50 tCO-2eq |

---

31  This threshold is in line with the limits stipulated for car emissions by kilometer according to EU climate action laws. For more information,
read "CO2 emission performance standards for cars and vans".
32  This is the value reported in "Estimating the Carbon Footprint of BLOOM, a 176B Parameter Language Model".

The following minimal ethical requirements tie normative recommendations to implementable practices. All requirements are general and should be employed regardless of the risk category of a system.

## Requirement SU1: Tracking of Environmental Markers

Tracking environmental markers such as CO2 emissions, energy use, and water consumption is essential in assessing and mitigating the ecological impacts of AI technologies. These markers serve as vital indicators of the sustainability of AI deployment. Understanding its environmental footprint becomes essential for responsible development and decision-making regarding the use and development of AI systems. Hence, tracking environmental markers is a minimal ethical requirement that should be stimulated across applications, independent of their risk.[33]

For example, monitoring CO2 emissions helps quantify the carbon footprint associated with AI infrastructure and operations, guiding efforts towards reducing greenhouse gas emissions. Likewise, tracking energy use provides insights into the efficiency of AI algorithms and hardware, facilitating optimizations to minimize energy consumption [38].[34]Meanwhile, monitoring water consumption is crucial for assessing AI technologies' indirect environmental impacts and effects on planetary boundaries. However, while there are straightforward mechanisms and methodologies to measure energy consumption and carbon emissions of hardware and software use, more ready-to-use tools are needed to allow stakeholders to track other ecological markers, like water usage, among other resources tied to developing some types of AI systems.

Hence, since carbon emissions do not account for factors like social impacts, legality, and rebound effects [200], other measures are required to track ecological markings. Hence, another way to track environmental markers is through lifecycle assessments (LCA). LCAs provide a holistic approach to evaluating the environmental impacts of AI technologies throughout their entire lifecycle, from raw material extraction to manufacturing, usage, and disposal [142, 24, 147]. Hence, the documentation of factors beyond carbon emissions, such as water usage, resource depletion, and waste generation, should also be considered a minimal ethical requirement. In short, LCAs can facilitate informed decision-making by identifying opportunities for eco-friendly design choices and resource-efficient practices.[35]

Lastly, the results of such tracking efforts should also be made public through comprehensive reporting, especially for applications regarded as high-risk. Reporting on ecological impacts allows for transparency and accountability in AI development and deployment, fostering trust among stakeholders and the public. It also enables policymakers, businesses, and consumers to make informed decisions, guiding efforts toward more sustainable practices and technologies.

## Requirement SU2: Sustainable and Open Development

For stakeholders seeking sustainable development practices within AI development, embracing openness and resource efficiency is imperative. Hence, when technological development is tied to the significant expense of resources (especially in high-risk settings), a shift towards sustainable and open development methodologies should be considered a viable approach to the starting point. In other words, instead of constantly reinventing the wheel, repurposing and building upon already established frameworks and infrastructure can significantly reduce the environmental footprint associated with AI development. For example, much of the development on applied language modeling for text generation applications is rooted in open models like BERT [49], RoBERTa [121], Mistral [96], and many others. These were developed once and reused/repurposed thousands of times, which can also be said for many other foundation models released with open licenses [179, 120, 225].

Open access technology, especially when requiring significant building resources, presents an opportunity for leveraging existing tech through recycling and expansion. This approach conserves resources and promotes collaboration and knowledge sharing within the AI community. Moreover, by embracing open development practices, barriers to entry are lowered, allowing for a more inclusive and diverse participation in AI innovation. This fosters a sustainable and socially responsible culture of innovation, addressing the broader ethical considerations inherent in AI development.[36]

---

33 To learn more about estimating environmental markers like CO2 and energy use, we recommend the "Energy Usage Reports: Environmental awareness as part of algorithmic accountability", illustrated in the methodology page of CodeCarbon's documentation.

34 You can use tools like CodeCarbon and Eco2AI to track energy use and carbon emissions of AI experiments and any demanding computational process.

35 Conducting LCAs for AI technologies poses several challenges, including data availability, methodological complexities, and uncertainty in the evaluation process. To help readers envision their own LCA reports, examples we mention are Apple's and Nvidia sustainability reports, and the "EcoChain's Life Cycle Assessment (LCA) – Complete Beginner's Guide".

36 Famous deep learning frameworks make available to developers a series of open-source foundations for downstream use in applications. Examples like PyTorch Hub, Hugging Face, Timm, TensorFlow, and Kaggle are only a few main examples where developers can find ready-to-use foundations that do not require expensive pre-training to be redone.

## Requirement SU3: Sustainable and Efficient Development

Modern state-of-the-art AI systems are usually a byproduct of a paradigm that requires much computing to succeed in areas where other paradigms have failed [194]. Hence, optimizing algorithms and models used and the hardware infrastructure to support it, are imperative to the sustainable development of AI [192, 45]. For example, energy efficiency in computational systems is paramount to foster sustainability, especially for applications categorized as high-risk (i.e., billion-parameter-sized neural networks), where design choices can heavily influence the resource consumption tied to their development. These choices extend to selecting components such as model architecture, training methodology, and hardware selection, which, when done wisely, are pivotal in optimizing resource consumption.

Regarding model architecture, certain design choices and algorithmic implementations can significantly impact the computation and, thus, require energy consumption for things like model training, evaluation, and inference [33, 140]. For instance, utilizing depthwise separable convolutions [34] instead of standard convolutional layers in computer vision tasks can substantially decrease computation requirements without severely compromising performance [84, 127]. Similarly, in language models, attention mechanisms pose a computational bottleneck, given the quadratic complexity growth that scales with sequence length [205], which can be alleviated through alternatives such as attention-free models [160, 78] or more efficient implementations, like FlashAttention [47, 46], group query attention [5], or sliding window attention [17], among others [216, 51].[37]

Dataset size also affects resource consumption. Scaling laws provide insights into the optimal dataset size relative to the model's parameters for a fixed budget. For instance, the Chinchilla scaling laws suggest maintaining a 20:1 tokens-parameter ratio for language models under a fixed compute budget [88]. Under this constraint, one can still achieve performance while also stipulating a limit to the amount of computing to be spent.[38]

Hardware selection is also crucial in optimizing energy consumption. State-of-the-art GPUs, though costly, significantly accelerate training runs and minimize long-term costs.[39] At the same time, efficient training techniques further enhance energy efficiency by optimizing model parameters without the resource-intensive nature of traditional methods. For example, Parameter Efficient Fine-Tuning (PEFT) methods [134], such as Low-rank adaptation (LoRA) [90], n-bit quantization [118, 209], memory efficient optimizers [186, 224], and Quantized low-rank adaptation (QLoRA) [48], are all methods that can make training and using large models much more efficient and less resource hungry.[40]

Ultimately, optimizing efficiency in computational systems involves strategic design choices in model architecture, dataset size considerations, and hardware selection, complemented by efficient training and inference techniques to minimize resource consumption. Adopting these measures is a minimal requirement for sustainable development, especially for high-risk AI systems and applications.

## Requirement SU4: Offsetting Policies

In pursuing sustainable and responsible AI development, offsetting policies emerge as a crucial requirement, particularly for high-risk applications. In the context of environmental sustainability, offsetting refers to counterbalancing or compensating for the negative ecological impacts of an activity by undertaking additional actions that result in positive environmental outcomes. Offsetting aims to achieve a net-zero or even a net-positive environmental impact, thereby attempting to mitigate the overall ecological footprint of a particular activity or project [126, 92]. One example is carbon offsetting. Carbon offsetting involves investing in projects or initiatives that reduce or capture carbon dioxide (CO2) emissions equivalent to those generated by a particular activity. For instance, if a company's operations produce a certain amount of CO2 emission, it can purchase carbon credits to fund projects such as reforestation, renewable energy development, or methane capture from landfills.

Offsetting policies provide an avenue to address the inherent trade-offs between technological progress and environmental sustainability. By quantifying the ecological cost of AI development and implementation (SU1), stakeholders can put in place strategies to neutralize or minimize these impacts. Additionally, offsetting policies promote accountability and transparency within the AI community. By requiring developers and organizations to assess and disclose the environmental implications of their projects, these policies foster a culture of environmental stewardship and responsible innovation.[41]

---

37 You can find several techniques to improve memory footprint and overall efficiency in training LLMs here, here, and here. For training improvements in diffusion models, check these implementations: MaskDiT and Patch Diffusion.

38 Here, you can calculate training and inference estimations for the resources required to train large models according to scaling laws.

39 The Deep Learning GPU Benchmark compares GPUs in terms of their latency regarding training, inference, and complexity of the task under consideration. Stakeholders can use these results, as well as the specifications of the intended hardware, to make informed decisions regarding how to optimize energy use.

40 Tools like Optimum provide several performance optimization tools, in line with the suggestions of this paragraph, to train and run models on targeted hardware with maximum efficiency.

41 You can learn more about Europe's carbon offsetting policies in the EU Emissions Trading System (EU ETS).

# 10 Transparency and Explainability (T)

Transparency, taking the definition provided by the AI Act and the analysis of 200 ethical guidelines [41] as our basis, encompasses the principle that the development and utilization of AI technologies should be open and understandable to all stakeholders. It makes information regarding the organization's AI practices and algorithms accessible and understandable for experts and non-experts alike. This principle necessitates that AI systems are designed and implemented to facilitate traceability and explainability, ensuring that humans are aware when interacting with AI systems. Additionally, transparency entails providing users with clear insights into the capabilities and limitations of AI systems and informing affected individuals about their rights concerning AI-generated decisions or interactions. Ultimately, transparency promotes accountability, trust, and ethical use of AI technologies within society.

From a practical perspective, transparency is a principle tied directly to the fields that seek to combat model opacity, like explainable AI (XAI) and mechanistic interpretability (MechInterp) [146, 210]. While XAI and MechInterp share the overarching goal of enhancing AI systems' intelligibility (especially machine learning models), they exhibit distinct methodologies and emphases. XAI primarily focuses on developing techniques to elucidate the outputs of opaque systems in an empirical fashion, where input-output relations are mapped in an observational fashion, e.g., input X generates output Y and components x and x are the major contributors for output Y, particularly those involving systems that are not hand-coded but developed by a learning paradigm. Techniques within the XAI paradigm include feature importance analysis [156], and model-agnostic approaches such as LIME (Local Interpretable Model-agnostic Explanations) and SHAP (SHapley Additive exPlanations) [175, 130].

In contrast, MechInterp centers on uncovering the underlying causal mechanisms governing observed phenomena. Unlike XAI, which operates primarily at the level of black-box models, MechInterp delves into the intrinsic structure and dynamics of systems, aiming to extract mechanistic insights that align with domain-specific knowledge. Techniques within MechInterp encompass mainly causal intervention methods like knowledge editing [139], circuit search [210], and reverse-engineering [151]. Despite their differences, XAI and MechInterp share several commonalities. Both recognize the importance of interpretability in fostering trust and facilitating human-machine collaboration, albeit from distinct vantage points. Furthermore, they confront similar challenges, such as the trade-off between model complexity and interpretability, the need to balance accuracy with comprehensibility, and the ethical considerations surrounding using interpretable models in high-stakes applications [180].

Other practices, which we could categorize as non-technical (do not involve the direct exploration of a system), are also recognized as practices in explainable and interpretable AI. This entails procedures like documentation and reporting, be that of general information and risk assessments [143] or more specific metrics tracked during development (e.g., carbon emissions) [111], improvement of AI literacy [123], and independent audits and reviews [61].

To help organizations prospect the level of risk regarding their system or application, we propose the following criteria, in line with the four main categories of risk identified in the EU AI Act (examples in Table 5):

1. **Minimal:**
   Systems that present a minimal risk regarding the principle of transparency are those in which their inputs and outputs are not directly involved with human beings in any critical way. These systems are the ones in which organizations are more lenient in using black box models, given that their inner workings, as long as they are reliable and robust, do not require our complete understanding.

2. **Limited:**
   Systems that present a limited risk regarding the principle of transparency are those whose inputs and outputs are directly involved with human beings, but their malfunction would only generate little harm. These systems are the ones in which organizations if chosen to utilize black box models, should provide results that can prove a minimal level of interpretability and understanding of their system.

3. **High:**
   Systems that present a high risk regarding the principle of transparency are those whose inputs and outputs are directly involved with human beings, and their malfunction would generate significant harm. These systems are the ones in which organizations should refrain from using black box models, given that such systems are tied to critical infrastructure (e.g., transport, healthcare, law enforcement, etc.).

4. **Unacceptable:**
   Systems that present unacceptable risk regarding the principle of transparency are those whose inputs and outputs are directly involved with human beings in a way that their malfunction would infringe on their human rights. Organizations should refrain from developing such systems or terminating them if already deployed.

Table 5: Examples of application areas and their categories of risk (Transparency)

| | |
|---|---|
| **Minimal:** | Spam Filters<br>Industrial Automation<br>Agriculture Automation |
| **Limited:** | AI Assistants<br>Automated Advertising<br>Recommendation Systems |
| **High:** | Legal Automation<br>Medical Diagnostics<br>Autonomous Vehicle |
| **Unacceptable:** | Social Scoring Systems<br>Automated Phishing<br>Remote Biometric Identification |

The following minimal ethical requirements tie normative recommendations to implementable practices. All requirements are general and should be employed regardless of the risk category of a system.

## Requirement T1: Explainable and Understandable Outcomes

The more risk involved in using an AI application or system, the more efforts should be made to explain its workings. Hence, systems regarded as low risks (e.g., spam filters) have the most leniency regarding these requirements. In contrast, in high-risk scenarios, stakeholders are required to invest heavily in XAI and MechInterp. By understandable explanations, we mean that meaningful attributions, correlations, and causal relations can be achieved properly and, when possible, a coherent narrative can be constructed, always keeping in mind that explanations should also have differential levels of understandability, from the expert to the layperson.

Hence, it is recommended that tools for model exploration be employed to produce such results.[42] For most applications, be that in computer vision, natural language processing, classification, or forecasting, there are methods, like LIME [175], SHAP [130], and tools, like DALEX [19], CAPTUM [108], and ALIBI [107], to aid stakeholders involved in this explanatory step to achieve this goal, remembering again that these explanations should always be tailored to the audience for which they are intended.

Tailored explanations for AI systems are crucial, especially in high-risk scenarios, to foster trust and accountability. Leveraging tools for model exploration and XAI techniques ensures transparent and understandable explanations, catering to diverse audiences and enhancing public confidence in AI systems.

## Requirement T2: Dataset Documentation

Data is a foundational element that dictates the behavior and efficacy of an AI system, making transparency regarding its characteristics paramount. Hence, a minimal ethical requirement for transparency and explainability is the documentation of datasets.

Dataset cards [167] serve as an instrument for illuminating the intricacies of the data employed in AI development. These cards encapsulate critical details about the dataset's composition, including its size, diversity, and provenance. By delineating the data collection methodology, annotation procedures, and potential biases within the dataset, stakeholders can gain insights into the underlying factors shaping the AI system's performance and outputs. Moreover, dataset cards facilitate understanding the contextual nuances surrounding the data, empowering researchers and practitioners to discern the implications of utilizing specific datasets in AI model training.[43]

Ultimately, dataset documentation contributes to establishing a more robust and accountable ecosystem wherein transparent and well-documented data sources underpin data-driven insights.

## Requirement T3: Model Reporting

To enable all AI stakeholders to perform their designated roles, the system's operations, intended usage, out-of-scope usage, performance, limitations, and risks should be documented clearly. Such documentation can aid in situations of failure or accidents, helping stakeholders set the scope of their responsibilities and liability.

This requirement can be achieved via model reporting [143]. Model reporting facilitates documentation regarding the AI system's operation by providing detailed insights into a given AI system, much like informational leaflets accompany pharmaceuticals. Through model cards, developers can disclose crucial information such as the architecture employed, training data sources, preprocessing techniques, and model performance metrics. Additionally, explanations regarding potential biases, limitations, and ethical considerations are elucidated, offering a comprehensive understanding of the AI system's functioning. This transparency fosters trust and accountability and enables users to make informed decisions about deploying and utilizing AI technology. Model cards are encouraged in many public repositories of open-source AI models (e.g., Hugging Face and GitHub). At the same time, such documents almost always accompany major releases of AI models.[44]

Users of AI products should be educated to request and scrutinize such documentation. At the same time, developers should be encouraged to create such documents plainly and understandably as a standard practice of their profession.

## Requirement T4: Risk Assessment in Black Box Scenarios

Black box models are characterized by complex internal workings often opaque to human understanding (e.g., billion parameter-sized neural networks). While these models can provide high performance in specific applications, they pose significant challenges for interpretability. Unlike more shallow linear models, where the relationship between input variables and output can be more easily understood, black box models obscure the logic behind their behavior, making it difficult for stakeholders to trust or interpret their decisions. This lack of interpretability raises concerns, particularly in high-risk settings where the consequences of errors can be severe.

Due to the inherent risks associated with black box models, as already recommended by prominent figures in the field [180], it's prudent to avoid their use in high-stakes environments whenever possible. In settings where human lives, financial stability, or ethical considerations are at stake, relying solely on opaque models can be considered ethically irresponsible. Instead, transparent models or approaches prioritizing interpretability should be favored.

---

42 For those technically inclined readers, we recommend "The Building Blocks of Interpretability" as a gentle introduction to the many techniques one could employ.

43 For more information on how to do this, have a look at "The Data Cards Playbook".

44 Example: Llama 2. You can use this application to fill up your model report.

By opting for models that offer insights into their decision-making process, stakeholders can better understand, validate, and potentially mitigate the risks associated with model errors.

In cases where black box models are unavoidable due to their superior performance or lack of viable alternatives, human moderation becomes essential. Implementing human oversight and control mechanisms can help mitigate the risks associated with these models. Hence, under such circumstances, moderators should monitor the behavior of black box models and intervene when necessary. Additionally, ongoing evaluation and auditing can help ensure that the models behave as intended and do not exhibit harmful biases or errors (e.g., red teaming [68]). While human moderation adds complexity and costs to deploying black box models, it is a crucial safeguard in mitigating their potential risks in high-risk settings.

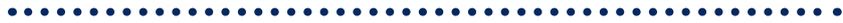

AI requirements emphasize explainable outcomes, dataset documentation, model transparency, avoiding black box models in high-risk scenarios, and human oversight to ensure safety, trust, and accountability.

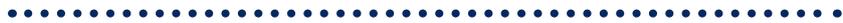

# 11 Truthfulness (TR)

The principle of truthfulness has emerged as a new matter of consideration for the field, mainly propelled by the recent strides in generative AI technologies [179, 86, 168, 3]. As these advancements enable AI systems to produce content that closely resembles, and in some cases, indistinguishably mirrors, human-generated content, the ethical imperative surrounding truthfulness has acquired newfound significance, which was a purely human problem until recently. This new preoccupation reflects a critical juncture wherein the traditional boundaries of truth, authenticity, and trustworthiness are redefined in light of AI's newly found capabilities. "Truth" because of how closely AI-generated data mimics human-generated content. "Authenticity" due to our need to certify the genuineness and originality of certain types of content. And "Trustworthiness" because, in simple terms, there can be no trust where there is no truth. Hence, much debate surrounds the issues related to the question, "What does it mean for society when the grain of truth becomes a harder-to-find spec in a sea of artificially generated content?".

However, what are truthfulness, falsehoods, and lies, and how can we define them in the context of generative AI? Firstly, truthfulness is only a consideration when discussing systems that generate content, like text, images, audio, or videos. Hence, we should not confound "model incompetence" or "mistakes" with falsehood. Questions related to how accurate a system is are better defined and dealt with when working with principles like safety, reliability, and robustness. Therefore, a system that should predict class "toxic" for a piece of harmful text but ends up outputting "neutral" is not "lying" or outputting falsehood. It simply is an inaccurate system.

At the same time, to define an AI-generated falsehood we must first recognize that for this condition of untruthfulness to hold, such systems are not required to be knowledgeable [58].[45] In other words, if a researcher trains a language model only on flat earth literature, and while prompted to answer "What is the shape of the Earth?" it outputs "a pancake", this would not be a case of falsehood, but model incompetence. At the same time, if a language model trained with the best current literature on physics outputs a statement later disproved by advances beyond its training cut-off, that is also not a case of falsehood but, again, of incompetence or, we might even say, ignorance.

If we agree upon these conditions, we can define AI falsehoods as cases where a generative model, which possesses access to the ground truth,[46] be that contained in its knowledge base or training data, generates a piece of content untied to that fact.[47] Another aspect that should also be considered an issue regarding truthfulness is when human actors use generative models to generate disinformation intentionally, like, for example, in the use of deep fakes to damage the reputation of individuals [131, 135]. Hence, either in the generation of misinformation or disinformation,[48] except for cases that can be attributed to model incompetence, stakeholders should promote measures that seek to mitigate the generation of AI-empowered falsehoods.

Even though the principle of truthfulness is still not a recurring theme in many published AI Guidelines [41],[49] there are many reasons why the proliferation of AI-empowered misinformation and disinformation should be a concern worth addressing. Given the impact such systems may have on society (e.g., undermining electoral processes [32]) and its individuals (e.g., reputational damage and identity theft [22]), following minimal requirements for truthful AI becomes necessary to promote trustworthy AI. Concerning the liability and consequences that should be enforced in the case of actors trying to spread falsehoods with AI intentionally, there remains to be seen how the AI Act will be implemented in practice (e.g., what sanctions are to be imposed on those who act in an adversarial fashion?).

To help organizations prospect the level of risk regarding their system or application, we propose the following criteria, in line with the four main categories of risk identified in the EU AI Act (examples in Table 6):

**1. Minimal:**

Systems with minimal risk regarding the principle of truthfulness are those whose output does not require factual grounding. In other words, those are applications tied to creative and artistic work. Also, minimal-risk systems are those whose outputs can be recognized as artificial without significant effort.

**2. Limited:**

Systems with a limited risk regarding the principle of truthfulness are those whose output requires considerable factual grounding. These systems and applications are usually tied to activities that require producing empirically, historically, and socially agreed-upon knowledge.

---

45 For more information on the matter, we recommend "Truthful AI: Developing and governing AI that does not lie".

46 We can define ground truth information known to be genuine, provided by direct observation and measurement (e.g., Earth is not flat). Also, we can describe it as information agreed upon by the majority, as is generally the case in matters in the humanities (e.g., Slavery is a morally abhorrent act).

47 The question regarding information being wrong due to human incompetence is, again, beyond the scope of this principle in the context of AI and this work. At the same time, when we speak of intentional falsehoods (i.e., disinformation), the intentionality, or cause, behind the act is the human intention. We are not considering the possibility of AI systems possessing or exhibiting their "own intentionality".

48 We can define misinformation as any incorrect or misleading information (it does not require an intentional component). In contrast, disinformation is false information deliberately spread to deceive people (it requires intentionality).

49 This is probably because many of these guidelines and ethical charters preceded the emergence of highly-skilled generative models [169, 25].

**3.** High:

Systems that present a high risk regarding the principle of truthfulness are those whose output requires considerable factual grounding and produce outputs that can only be distinguished from human-generated content with significant effort.

**4.** Unacceptable:

Systems that present unacceptable risks regarding the principle of truthfulness are those whose output can be indistinguishable from human-generated content. Additionally, they are those that, besides being used in domains that require considerable factual grounding, through them, be by the absence of guardrails or mitigating strategies, users can intentionally deceive or manipulate others by utilizing such systems to present false information as truth, causing harm, or undermining trust in information sources in a scalable fashion. Organizations should refrain from developing such systems or terminating them if already deployed.

Table 6: Examples of application areas and their categories of risk (Truthfulness)

| | |
|---|---|
| **Minimal:** | Art Generation<br>Creative Writing Assistant |
| **Limited:** | Coding Assistants<br>Search-Engine Assistants |
| **High:** | Photo Realistic Image Editing<br>Photo Realistic Image Generation |
| **Unacceptable:** | Deep Fakes<br>Automated Phishing |

The following minimal ethical requirements tie normative recommendations to implementable practices. All requirements are general and should be employed regardless of the risk category of a system.

## Requirement TR1: Disclosure and Watermarking

Nowadays, AI-driven technologies are seamlessly integrated into various facets of our daily lives, often blurring the lines between humans and machines. For instance, in video games, AI-generated conversations simulate human-like exchanges, enriching the gaming experience by creating dynamic and immersive environments. However, amid this technological marvel, it becomes imperative to underscore the significance of disclosure in AI applications, particularly those that mimic human behavior, such as text-to-image and language models, to avoid deceiving or deluding less informed stakeholders.[50]

Disclosure in these types of AI applications[51] is paramount for several reasons. Firstly, it ensures transparency and honesty in user interactions, establishing trust between users and the technology. Also, communicating to users that they are interacting with an AI system, not a human, helps manage expectations and prevents potential misunderstandings and deception. Additionally, disclosure allows users to understand the limitations of the technology they are engaging with, avoiding liability problems. For example, by warning users about potential issues like hallucinations and the risk of over-anthropomorphization inherent in AI systems, individuals can approach interactions with a critical mindset and make more informed decisions about the information they receive.

Here are some practical measures organizations can implement to promote open and honest disclosure in AI applications:

- Clear and prominent labeling should indicate when users interact with AI systems rather than human counterparts. This labeling can include visible notifications or disclaimers at the onset of interactions, ensuring users are aware of the AI's involvement.

- Organizations should provide accessible documentation detailing the capabilities and limitations of their AI systems. This documentation should outline potential issues such as hallucinations, verbosity, repetition, and biases, empowering users to make informed decisions during their interactions. Akin to one of the requirements for transparency (T3), model cards [143] and similar reporting methods (Terms of Use) can complement this practice.

- Organizations should also establish channels for feedback and complaints, which, besides fostering open communication and accountability, ensure individuals can contact a human mediator if needed (i.e., inserting a human-in-the-loop).

Another way to promote truthfulness and disclosure is via watermarking. Watermarking is a technique extensively employed in cryptography and steganography that offers a promising avenue for enhancing veracity in generative AI applications. While steganography aims to hide secret information within digital media without changing the visible appearance, watermarking aims to embed information that verifies the owner or the authenticity of that media [98, 187, 211, 60]. In the context of generative AI, this methodology can be adapted to imbue AI-generated outputs with traceable markers, enabling the creation of digital fingerprints that allow stakeholders to track and validate AI-generated artifacts [105, 122]. By integrating watermarking mechanisms into the fabric of generative AI systems, researchers and stakeholders can address concerns related to misinformation, intellectual property infringement, and overall trustworthiness.

To further aid in watermarking efforts, adding identifiable metadata to AI-generated content should also be required in high-risk settings. For example, C2PA metadata is an open technical standard that allows organizations to embed metadata in media to verify its origin [196, 65, 10]. This extra step further helps create a causal link between AI-generated media and all stakeholders involved.[52]

---

## Requirement TR2: Factual Grounding

In applications necessitating factual groundedness, model hallucination emerges as a prominent concern within generative AI systems. We can define hallucination in AI as a generated output containing false or misleading information presented as fact like incorrect, nonsensical, or unreal text [137, 95]. This phenomenon poses significant challenges, particularly in domains where precision and reliability are paramount, such as education and journalistic reporting. The occurrence of hallucinations not only undermines the integrity of generated content but also jeopardizes user trust and confidence in AI-generated outputs. Furthermore, in contexts where decision-making relies on the information provided by AI systems, hallucinations can lead to erroneous conclusions and adverse outcomes.

One promising approach to combat model hallucination in generative AI systems is grounding them in trusted and curated data sources, such as external knowledge pools. This strategy, where generative AI systems are coupled to retrieval systems that search and find relevant information, is called retrieval augmented generation (RAG) [116, 70]. Retrieval generation allows AI systems to, much like humans do, increase their knowledge and application scope by accessing large pools of curated information and, through their in-context learning skills, integrate the retrieved data into their outputs. This approach can help reduce the likelihood of hallucinations and improve the quality of generated outputs [188]. This hybrid approach offers a promising avenue for combating hallucinations. RAG should be utilized in applications that require factual groundedness, given we cannot (currently) comprehend and audit how large generative models store and retrieve information from inscrutable matrices of floating point numbers.

While grounding generative AI in trusted data sources and adopting retrieval augmented generation techniques are crucial in mitigating model hallucination, rigorously evaluating such systems' groundedness and factuality remains imperative. Hence, like in safety and robustness (SR1), generative systems (augmented by retrievers or not) need to be evaluated on benchmarks set to assess the propensity of a model to generate falsehoods.[53] This way, stakeholders can use benchmark evaluations to determine further how truthful a system is.

## Requirement TR3: Fact Checking and Guardrails

When discussing fact-checking, we mean verifying the accuracy and truthfulness of claims, statements, or information circulating within various media platforms [6, 76, 203]. When humans perform these activities, fact-checking involves rigorous investigation, comparison with credible sources, and analysis of evidence to determine the validity of the information in question. In applications involving high-risk settings, human moderators should be used to help flag and catalog AI-generated falsehoods, much like is already done in several media platforms [199, 71]. However, as already pointed out by the literature [7, 23], human oversight, when not augmented, has difficulties in accompanying the demands of large-scale automation. For this reason, creating automated forms of automated fact-checking guardrails should also be seen as a requirement.

In the context of fact-checking, guardrails refer to a set of technical constraints designed to ensure the truthfulness of a system. For example, automated fact-checking systems can help expedite and scale human moderation while also flagging or blocking content that is deemed untruthful. Hence, to improve the scalability of human oversight, AI-empowered guardrails can further help improve the condition of a system. Using other AI tools to help us guard other

AI systems is a common practice in the field [173, 53, 171], and stakeholders should employ them in the establishment of fail-safe procedures, especially in applications involving a limited, or high, risk.[54]

By leveraging technological advancements to aid human control, fact-checking initiatives can effectively counter the proliferation of misleading content, safeguarding individuals' power to make informed decisions and fostering a healthier information ecosystem between humans and AI.

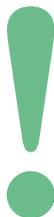

Transparency and accuracy in AI systems are essential for building user trust, necessitating clear disclosure of AI usage, grounding in reliable data sources, and implementing automated fact-checking alongside human moderation to prevent misleading content.

---

53 For modalities involving language, benchmarks like TruthfulQA [119] and FactCheckQA [14] can help determine the truthfulness of a system. Meanwhile, ArtiFact and [170] DeepfakeBench [219] can help in dealing with applications involving computer vision. Lastly, DEEP-VOICE [20] can aid in detecting and evaluating AI-generated speech.

54 AI-powered guardrails can be used to, for example, identify and debunk fabricated narratives (Google's FactCheck Tools), improve factual consistency (LangCheck), detect factual inaccuracy (UpTrain), and overall misinformation detection (FacTool).

# 12 Concluding Remarks

Artificial intelligence has seamlessly woven itself into the fabric of our daily lives, from personalized recommendations on streaming platforms to intelligent assistants that fulfill our needs. However, its integration has sparked profound questions about the boundaries of technology. As AI permeates various sectors, concerns about privacy, autonomy, and many other ethical matters arise. Hence, as we navigate this era of unprecedented technological advancement, it becomes imperative to assess the implications of AI integration critically and actively shape its evolution to align with societal values and aspirations.

Regulating and certifying AI ensures this technology's trustworthy and ethical deployment. However, stakeholders need clear criteria and guidelines to assess whether development is made with societal values in mind. To aid on this front, we presented two sets of ethical and practical guidance to the community in this document. While our overall ethical requirements (O1 – O6) represent general and holistic values that should serve as guiding foundations for trustworthy AI development, our value-specific requirements (F, P, SR, SU, T, TR) deliver more pragmatic and implementational requirements and procedures, alongside with possible criteria of evaluation in line with the EU AI Act risk-category system.

At the same time, certifying AI systems requires assessing that minimal ethical requirements are fulfilled and dealt with in practice. Nevertheless, bridging the principles-practice gap is an ongoing area of research and exploration that poses a considerable challenge to applied ethics. In this work, we highlight and suggest tools that could tackle specific challenges related to trustworthy AI development to the reader. Yet, we would like to point readers to other resources that approach this aspect even more thoroughly. The Catalog of Tools and Metrics from the OECD,[55] which currently harbors more than 700 practical implementations for trustworthy AI development, be that procedural tools[56] to aid in the lifecycle of AI systems or educational tools to help improve humanistic and ethical practices in IT-related fields.

Finally, we highlight that this is an ongoing project, bound to be changed and adapted as the field progresses. Regardless, we hope our efforts can be used, expanded, and built upon by the community in search of an ever more trustworthy AI.

## Acknowledgments

We thank Sergio Genovesi and Marta Cassina for their contributions to the initial draft of the milestone, Chelsea Haramia, Was Rahman, and Christiane Schäfer for their comments in our research group meeting, Was Rahman for comments during the milestone's editing process and Sophia Falk for comments on the whitepaper draft.

---

55 The OECD's "Catalogue of Tools & Metrics for Trustworthy AI", for example, presents tools and metrics designed to help AI actors develop and use trustworthy AI systems and applications that respect human rights and are fair, transparent, explainable, robust, secure, and safe.
56 As an example, we can mention Ethical Problem Solving by Nicholas Kluge Corrêa et al., a framework to promote the development of safe and ethical artificial intelligence via algorithmic impact assessment tools (e.g. the Teeny-Tiny Castle), and a recommendation methodology that culminates in an extensive developmental toolbox.

# 13 Ethical requirements at a glance

### Overall Ethical Requirements (O)

- ☐ O1: Human Oversight and Control
- ☐ O2: Impact Mitigation and Robust Deployment
- ☐ O3: Open Development and Transparent Reporting
- ☐ O4: Privacy, Diversity and Accessibility in, by and for Design
- ☐ O5: Ethical Stakeholder Inclusion and Co-creation
- ☐ O6: Democracy and Human Rights

### Fairness (F)

- ☐ F1: Bias Analysis
- ☐ F2: Open Disclosure
- ☐ F3: Rectify or Mitigate Biased Outcomes
- ☐ F4: Human Moderation

### Privacy and Data Protection (P)

- ☐ P1: Purpose Limitation and Data Minimization
- ☐ P2: User Control and Information
- ☐ P3: Data Protection by Design and Default
- ☐ P4: Anonymization and Pseudonymization
- ☐ P5: Data Protection Impact Assessment
- ☐ P6: Codes of Conduct and Privacy Management

### Safety and Robustness (SR)

- ☐ SR1: Robust and Open Evaluation
- ☐ SR2: Red Teaming and Broader Impact Analysis
- ☐ SR3: Human Oversight

### Sustainability (SU)

- ☐ SU1: Tracking of Environmental Markers
- ☐ SU2: Sustainable and Open Development
- ☐ SU3: Sustainable and Efficient Development
- ☐ SU4: Offsetting Policies

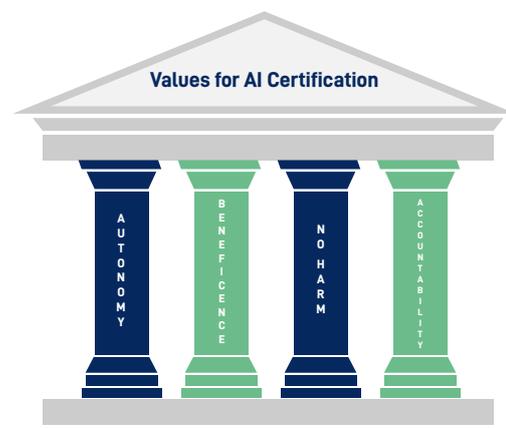